\documentclass[pre,aps,twocolumn]{revtex4}
\usepackage{epsfig,latexsym}
\usepackage[hypertex]{hyperref}

\begin{document}

\title{Universal statistical properties of poker tournaments}

\author{Cl\'ement Sire\footnote{{\it E-mail: {\it
clement@irsamc.ups-tlse.fr}; Web: www.lpt.ups-tlse.fr}}}

\affiliation{Laboratoire de Physique Th\'eorique - IRSAMC, Universit\'e
Paul Sabatier \& CNRS, Toulouse, France}

\begin{abstract}
We present a simple model of Texas hold'em poker tournaments which
retains the two main aspects of the game: \emph{i.\,}the minimal bet
grows exponentially with time; \emph{ii.\,}players have a finite
probability to bet all their money. The distribution of the fortunes of
players not yet eliminated is found
to be independent of time during most of the tournament, and
reproduces accurately data obtained from Internet tournaments and
world championship events. This model also makes the connection
between poker and the persistence problem widely studied in physics,
as well as some recent physical models of biological evolution, and
extreme value statistics.
\end{abstract}

\maketitle

\section{Introduction}

Physicists are now more then ever involved in the study of complex
systems which do not belong to the traditional realm of their
science. Finance (options theory,...) \cite{bouchaud}, human networks
(Internet, airports,...) \cite{barabasi}, the dynamics of biological
evolution \cite{krug,leadcs} and in general of competitive ``agents''
\cite{krap,red1,red2} are just a few examples of problems recently
addressed by statistical physicists.  However, many of these systems
are not isolated and are thus sometimes very difficult to describe
quantitatively: a financial model cannot predict the occurrence of
wars or natural disasters which certainly affect financial markets,
nor can it include the effect of all important external parameters
(China's GDP growth, German exports, Google's profit...). Rather,
these studies try to capture important qualitative features which,
interestingly, are sometimes {\it universal}. In this context,
universality means that large scale aspects of the real system are
properly reproduced by a simple model which only retains the main
relevant ingredients of the original physics. Adding further details
to the model does not affect these universal properties.

In the present work, we study a very human and playful activity:
poker tournaments. Although {\it a priori} governed by human laws
(bluff, prudence, aggressiveness...), we shall find that some of
their interesting properties can be quantitatively described. One of
the appealing aspects of a poker tournament lies in the obvious fact
that it is a truly {\it isolated system}, which is not affected by
any external phenomenon. Two famous mathematicians (\'Emile Borel
\cite{borel}, and later John von Neumann \cite{neumann}) contributed
to the science of poker. However, they concentrated on head-to-head
games, like their most recent followers \cite{ferg}, obtaining the
best strategy in terms of the value of the hand and the pot. To our
knowledge, the present work represents the first study of large
scale poker tournaments. Note however that in a recent work
\cite{red1}, the authors study head-to-head elimination tournaments
involving seeded competitors, and apply successfully their theory to
the US college basketball national championship.

In the following, we introduce a simple model which can be treated
analytically and which faithfully reproduces some properties of
Internet and live poker tournaments. Our main quantities of interest
are the distribution of the fortunes of surviving players, their decay
rate, the number of different players owning the biggest fortune
at any given time during the tournament (dubbed the ``chip leader''),
and the distribution of their fortune. Interestingly, the constraint
that a surviving player must keep a positive fortune relates poker
tournaments to the problem of persistence \cite{AB1,per1,SM}, and the
competitive nature of the game connects some of our results with
recent models of competing agents
\cite{krug,leadcs,krap,red1,red2}. In addition, the properties of the
chip leader display extreme value statistics, a phenomenon
observed in many physical systems \cite{red1,extreme,gumbel}.

In Section II, we define a stochastic model which retains the main
identified ingredients of poker tournaments: \emph{i.\,}the minimal
bet grows exponentially with time; \emph{ii.\,}players have a finite
probability $q$ to bet all their money. In Section III, we first solve
the corresponding model for $q=0$, which will allow us to make the
connection with the persistence problem widely studied by
physicists. In Section IV, we will show that $q$ must physically take
a specific value, and thus is not a free parameter. The results of the
model will compare favorably with actual data recorded from real
Internet poker tournaments and World Poker Tour main events. Finally,
in the last Section V, we will consider the statistical properties of
the chip leader. In particular, we will show the connection with
the ``leader problem'' arising in evolutionary biophysics, and the
field of extreme value statistics which has recently attracted a lot
of attention from physicists.

\section{A simple poker model}

Before addressing the basic rules of poker and the
resulting definition of our model, we wish to introduce some useful
poker terminology. In a real poker tournament, players first pay the
same entry fee or ``buy-in'' (from 1$\,$\$ to 25000$\,$\$) which is
converted in ``chips''.  Hence players are not betting actual money
but chips. The total number of chips of a player is called his
``stack''. At any time in the tournament, if a player decides to bet his
entire stack, it is said that he is going ``all-in''.

We now describe the main aspects of a Texas hold'em poker tournament,
currently the most popular form of poker. Initially, $N_0$ players sit
around tables accepting up to $\theta=10$ players. In real poker
tournaments, $N_0$ typically lies in the range $N_0\sim 10-10000$.  We
do not detail the precise rules of Texas hold'em poker, as we shall
see that their actual form is totally irrelevant provided that two
crucial ingredients of the game are kept:

$\bullet$ A tournament consists in a series of independent games or
``deals''. Before a deal starts, the two players next to the dealer
(\textit{i.e.} the player dealing the cards) post the minimal bet,
which is called the ``blind''. This term arises from the fact that
they bet before actually seeing their cards. The blinds also ensure
that there is some money in the pot to play for at the very start of
the game. The blind $b$ \emph{increases exponentially} with time, and
typically changes to the value 40\,\$, 60\,\$, 100\,\$, 200\,\$,
300\,\$, 400\,\$,... every 10-15 minutes on Internet tournaments,
hence being multiplied by a factor 10 every hour or so. We shall see
that the growth rate of the blind entirely controls the pace of a
tournament, a phenomenon observed in another context in
\cite{red2}. {Therefore, the fact that the blind grows exponentially
with time must be a major ingredient of any realistic model of poker.}

$\bullet$ The next players post their bets ($\geq b$) according to
their evaluation of the two cards they each receive. There are
subsequent rounds of betting following the successive draws of five
common cards. Ultimately, the betting player with the best hand of
five cards (selected from its two cards and the five common cards)
wins the pot. Most of the deals end up with a player
winning a small multiple of the blind. However, during certain deals,
two or more players can aggressively raise each other, so that they
finally bet a large fraction, if not all, of their chips. This can
happen when a player goes all-in, hence betting all his
chips. {Any se\-rious mo\-del of poker should take into account
the fact that players often bet a few blinds, but sometimes end up
betting all or a large fraction of their chips.}

Once a player loses all his chips, he is eliminated. During the
course of the tournament, some players may be redistributed to other
tables, in order to keep the number of tables minimum.

Retaining the two main ingredients mentioned above, we now define a
simple version of poker which turns out to describe
\emph{quantitatively the evolution of real poker tournaments}. The
$N_0$ initial players are distributed at tables with $\theta =10$
seats. They receive the same amount of chips $x_0\gg b_0$, where $b_0$
is the initial blind. The ratio $x_0/b_0$ is typically in the range
$50-100$ in actual poker tournaments.

$\bullet$ The players take turns at dealing. In the model, only the
player next to the dealer, dubbed the ``blinder'', posts the blind
bet. The blind increases exponentially with time as,
$b(t)=b_0\exp(t/t_0)$.

$\bullet$ The tables run in parallel. At each table, the players
receive one card, $c$, which is a random number uniformly distributed
between 0 and 1.

$\bullet$ We define a critical hand value $c_0\in[0,1]$. The
following players bet the value $b$ with probability $e(c)$, if
$0\leq c\leq c_0$. $e(c)$ is an evaluation function, whose details
will be immaterial. Intuitively, $e(c)$ should be an increasing
function of $c$, implying that a player will more often play good
hands than bad ones. We tried several forms of $e(c)$, obtaining the
same results. In our simulations, we choose $e(c)=c^n$, where $n$ is
the number of players having already bet, including the blinder. In
this case, $e(c)$ is simply the probability that $c$ is the best
card among $n+1$ random cards. This reflects the fact that a player
should be careful when playing bad hands if many players have
already bet. Determining the optimal evaluation function for a given
$\theta$, in the spirit of Borel's and von Neumann's analysis for
$\theta=2$, is a formidable task which is left for a future study
\cite{csnext}.

$\bullet$ The first player with a card $c>c_0$ goes all-in, so that
$q=1-c_0$ is the probability to go all-in. The next players
including the blinder can follow if their card is greater than
$c_0$, and fold otherwise. If a player with a card $c>c_0$ cannot
match the amount of chips of the first player all-in, he simply bets
all his chips, but can only expect to win this amount from each of
the other players going all-in.

$\bullet$ Finally, the betting player with the highest card wins
the pot and the blinder gets the blind back if nobody else bets. The
players left with no chips are eliminated, and after each deal,
certain players may be redistributed to other tables, in a process
ensuring that the number of tables remains minimum at all times and
that no table has less than $[N/\theta]$ players, where $[\bf{\cdot}]$
denotes the integer part function. After a deal is completed at all
tables, time is updated to $t+1$, and the next deal starts. This
process is repeated until only one player is left.

\section{Poker model without all-in processes}

Let us first consider the unrealistic case $q=0$. The amount of
chips or stack $x(t)$ of a given player evolves according to
$x(t+1)=x(t)+\varepsilon(t)b(t)$. The effective noise
$\varepsilon(t)$ should have zero average since all players are
considered equal and there is therefore no individual winning
strategy in the mathematical sense. $\varepsilon(t)$ is also
Markovian, since successive deals are uncorrelated. We define $\bar
x(t)$ as the statistical average of $x(t)$. If the typical value of
$x\sim\bar x(t)$ remains significantly bigger than the blind $b(t)$,
we can adopt a continuous time approach. Hence, the evolution of
$x(t)$ is that of a generalized Brownian walker:
\begin{equation}
\frac{dx}{dt}=\sigma b(t)\eta(t),
\end{equation}
where $\sigma^2=\bar{\varepsilon^2}$ is a constant of order unity,
and $\eta(t)$ is a $\delta$-correlated white noise. The number of
surviving players with $x$ chips, $P(x,t)$, evolves according to the
Fokker-Planck equation
\begin{equation}
\frac{\partial P}{\partial t}=\frac{\sigma^2b^2(t)}{2}
\frac{\partial^2 P}{\partial x^2},\label{fokker}
\end{equation}
with the absorbing boundary condition $P(x=0,t)=0$, and initial
condition $P(x,t=0)=\delta(x-x_0)$. This kind of problem arises
naturally in physics in the context of persistence, which is the
probability that a random process $x(t)$ never falls below a certain
level \cite{AB1,per1,SM}. Defining
\begin{equation}
\tau(t)=\frac{\sigma^2b_0^2t_0}{2}\left({\rm e}^{\frac{2t}{t_0}}-1\right),
\end{equation}
Eq.~(\ref{fokker}) can be solved by the method of images
\cite{per1}:
\begin{equation}
P(x,t)=\frac{N_0}{\sqrt{2\pi \tau(t)}}\left({\rm
e}^{-\frac{(x-x_0)^2}{2\tau(t)}}- {\rm
e}^{-\frac{(x+x_0)^2}{2\tau(t)}}\right).
\end{equation}
Note that the above result holds for any form of $b(t)$, provided
that one properly defines $\tau(t)=\sigma^2\int_0^{t} b^2(t')\,dt'$.

For large time (or $\tau\gg 1$), the distribution of chips becomes
scale invariant
\begin{equation}
P(x,t)=\frac{N(t)}{\bar x(t)} f\left(\frac{x}{\bar
x(t)}\right),
\end{equation}
where the density of surviving players is given by
\begin{equation}
\frac{N(t)}{N_0}=\frac{2x_0}{\sqrt{\pi t_0}\sigma b_0}
{\rm e}^{-\frac{t}{t_0}}.
\end{equation}
We find that the decay rate of the number of players is exactly
given by the growth rate of the blind, which thus controls the pace
of the tournament. The total duration of a tournament $t_{\rm f}$ is
typically
\begin{equation}
\frac{t_{\rm f}}{t_0}=\ln
(N_0)-\frac{1}{2}\ln(t_0)+\ln\left(\frac{x_0}{b_0} \right),
\end{equation}
which only grows logarithmically with the number of players and the
ratio ${x_0}/{b_0}$. The average stack is proportional to the blind
\begin{equation}
{\bar x(t)}=\frac{N_0}{N(t)}x_0=\frac{\sqrt{\pi t_0}\sigma}{2}b(t).
\end{equation}
When $t_0\gg 1$, this expression implies that ${\bar x(t)}/b(t)\gg
1$, hence validating the use of a continuous time approach. Finally,
we find that the normalized distribution of chips is given by the
Wigner distribution
\begin{equation}
f(X)=\frac{\pi }{2}X{\rm e}^{-\frac{\pi }{4}X^2},\quad F(X)=1-
{\rm e}^{-\frac{\pi }{4}X^2}\label{scal1},
\end{equation}
\begin{figure}
\psfig{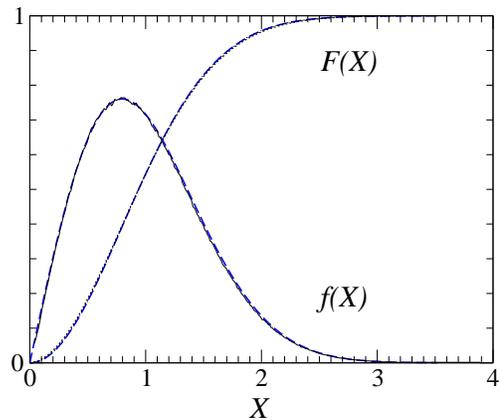}
\caption{\label{fig1} We plot the normalized distribution of chips $f(X)$ and
its cumulative sum $F(X)$ obtained from numerical simulations of our poker
model with $q=0$ (thin lines; $N_0=10000$, $t_0=2000$, $x_0/b_0=100$,
$10000$ ``tournaments'' played). These distributions are extracted at
times for which $N(t)/N_0=50$\%, 30\%, 10\%. The dashed lines
correspond to the exact solution, Eq.~(\ref{scal1}).}
\end{figure}
where $F(X)=\int_0^Xf(Y)\,dY$.  Equivalently, in the context of
persistence, $f$ is naturally found to be the first excited
eigenstate of the quantum harmonic oscillator \cite{per1}. The
scaling function $f$ is \emph{universal}, \emph{i.e.} independent of
all the microscopic parameters ($b_0$, $t_0$, $x_0$...). In
Fig.~\ref{fig1}, we plot the normalized distribution $f(X)=\bar
x(t)P(x,t)/N(t)$ and its cumulative sum $F(X)$ as a function of
$X=x/\bar x(t)$, as obtained from extensive numerical simulations of
the present poker model with $q=0$. We find a perfect data collapse
on the analytical result of Eq.~(\ref{scal1}).

\section{Poker model including all-in processes}

Let us now consider the more realistic case $q>0$ (or $c_0<1$).
\emph{A priori}, it seems that $q$ is a new parameter whose precise
value could dramatically affect the dynamics of the game. In
reality, $q$ must be intimately related to the decay rate $t_0^{-1}$
of the number of players, which is imposed by the exponential growth
of the blind. To see this, let us first compute the decay rate due
to the all-in processes. At a given table, and for small $q$, the
probability that an all-in process occurs is given by
\begin{equation}
P_{\rm all-in}=\frac{q^2\theta(\theta-1)}{2},
\end{equation}
where the factor $q^2$ is the probability that two players go
all-in, and ${\theta(\theta-1)}/{2}$ is the number of such pairs.
Expecting $q\ll 1$, we have neglected all-in processes involving
more than two players. During a two-player all-in process, there is
a probability $1/2$ that the losing player is the one with the
smallest number of chips (he is then eliminated). Cumulating the
results of the $N/\theta$ tables, we find the density decay rate due
to all-in processes
\begin{eqnarray}
\frac{dN}{dt}_{\rm all-in}&=&-\frac{1}{2}{\times}\frac{N}{\theta}
{\times} P_{\rm all-in}=-\frac{N}{t_{\rm all-in}},\\
t_{\rm all-in}&=&\frac{4}{q^2(\theta-1)}.
\end{eqnarray}
We now make the claim that the physically optimal choice for $t_{\rm
all-in}$, and hence for $q$, is such that the decay rate due to
all-in processes is equal to the one caused by the stack
fluctuations of order $b(t)$. Since the total decay rate will be
shown to remain equal to $t_0^{-1}$, $t_{\rm all-in}=2t_0$ should
hold, since inverse decay rates add up. If $t_{\rm all-in}<2t_0$, the
game is dominated by all-in processes and $x(t)$ can get rapidly
large compared to $b(t)$. The first player to go all-in is acting
foolishly and takes the risk of being eliminated just to win the
negligible blind. Inversely, if $t_{\rm all-in}>2t_0$, players
(especially those with a declining stack) would be foolish not to
make the most of the opportunity to double their chips by going
all-in. We expect that real poker players would, on average,
self-adjust their $q$ to its optimal value. Finally, we find that
$q$ is not a free parameter, but should take the physical value
\begin{equation}
q=\sqrt{\frac{2}{(\theta-1)t_0}}.\label{q}
\end{equation}

We now write the exact evolution equation for the number of
surviving players with $x$ chips, combining the effect of pots of
order $b$ and all-in processes
\begin{equation}
\frac{\partial P}{\partial t}=\frac{\sigma^2b^2}{2}
\frac{\partial^2 P}{\partial x^2}+\frac{2}{t_0}(K(P)-P),\label{fokker1}
\end{equation}
where the non linear all-in kernel $K$ is given by
\begin{eqnarray}
K(P)&=&\frac{1}{4}P(x/2)\int_{x/2}^{+\infty}\frac{P(y)}{N}\,dy\nonumber\\
     &+&\frac{1}{2}\int_0^{x/2}P(x-y)\frac{P(y)}{N}\,dy\nonumber\\
     &+&\frac{1}{2}\int_0^{+\infty}P(x+y)\frac{P(y)}{N}\,dy,
\label{kernel}
\end{eqnarray}
and where we have dropped the time variable argument for clarity. In
Eq.~(\ref{fokker1}), the factor ${2}/{t_0}=q^2(\theta-1)$ is simply
the rate of all-in processes involving the considered player,
without presuming the outcome of the event. In addition, the first
term of Eq.~(\ref{kernel}) describes processes where the considered
player has doubled his chips by winning against a player with more
chips than him. The second term corresponds to an all-in process
where the player has won against a player with less chips than him
(and has eliminated this player). Finally, the last term describes
the loss against a player with less chips than him (otherwise the
considered player is eliminated). Integrating Eq.~(\ref{kernel})
over $x$, we check that the probability to survive an all-in process
is $\frac{3}{4}$, the two first terms adding up to $\frac{1}{2}$.
Indeed, the player survives if he wins (with probability
$\frac{1}{2}$) or if he loses, but only against a player with less
chips (with probability $\frac{1}{4}$). We recover the decay rate
associated to all-in processes,
$\left(1-\frac{3}{4}\right){\times}\frac{2}{t_0}=t_{\rm all-in}^{-1}$.

We now look for a scaling solution of Eq.~(\ref{kernel}) of the form
\begin{equation}
P(x,t)=\frac{\lambda}{\hat x(t)^2}f\left(\frac{x}{\hat x(t)}\right),
\end{equation}
where the integral of $f$ is normalized to 1, so that
${N(t)}={\lambda}/{\hat x(t)}$. Plugging this \emph{ansatz} into
Eq.~(\ref{fokker1}), we find that one must have $\hat x(t)\sim b(t)$
for all the terms to scale in the same manner. Defining
\begin{equation}
\hat x(t)=\frac{\sqrt{t_0}\sigma b(t)}{2}\sim \bar x(t) \sim {\rm
e}^{\frac{t}{t_0}},
\end{equation}
and the scaling variable $X=x/\hat x(t)$, we
obtain the following integrodifferential equation for $f(X)$
\begin{eqnarray}
&f''(X)+X f'(X)+\frac{1}{2}f(X/2)\int_{X/2}^{+\infty}f(Y)\,dY\nonumber\\
&+\int_0^{X/2}f(X-Y)f(Y)\,dY\nonumber\\
&+\frac{1}{2}\int_0^{+\infty}f(X+Y)f(Y)\,dY=0,
\label{kernel2}
\end{eqnarray}
with the boundary condition $f(0)=0$. We did not succeed in solving
this equation analytically. However, the small and large $X$
behavior of $f(X)$ can be extracted from  Eq.~(\ref{kernel2}):
\begin{equation}
f(X)\mathop{\sim}\limits_{X \to 0} \frac{X}{2},\quad \, f(X)
\mathop{\sim}\limits_{X \to +\infty} 2\mu {\rm{e}}^{- \mu X}.\label{expdecay}
\end{equation}
Thus, when including all-in processes, the universal scaling
distribution decays more slowly than for $q=0$. Eq.~(\ref{kernel2})
can be easily solved numerically using a standard iteration scheme,
and we find $\mu\approx 1.562$.
\begin{figure}
\psfig{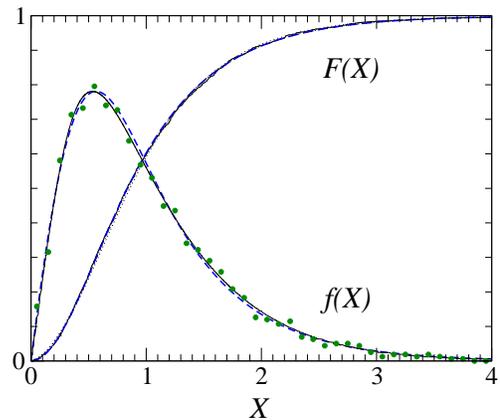}
\caption{\label{fig2} We plot the normalized distribution of chips
$f(X)$ and its cumulative sum $F(X)$ obtained from numerical
simulations of our poker model (thin dotted lines, $N_0=10000$,
$t_0=2000$, $x_0/b_0=100$, $10000$ ``tournaments''
played). These distributions are extracted at times for which
$N(t)/N_0=50$\%, 30\%, 10\%. The dashed lines correspond to the
numerical solution of the exact Eq.~(\ref{kernel2}). The data
recorded from 20 real poker tournaments (totalizing 1584 players still
in) are also plotted (full lines), and are found to agree
remarkably with the present theory. Note that $f(X)$ for real
tournaments was obtained by differentiating a fitting function to the
actual cumulative sum. We also plot the standard but noisier bin plot
of the distribution of chips in real poker tournaments (circles).}
\end{figure}

\begin{figure}
\hspace{-0.6cm}\psfig{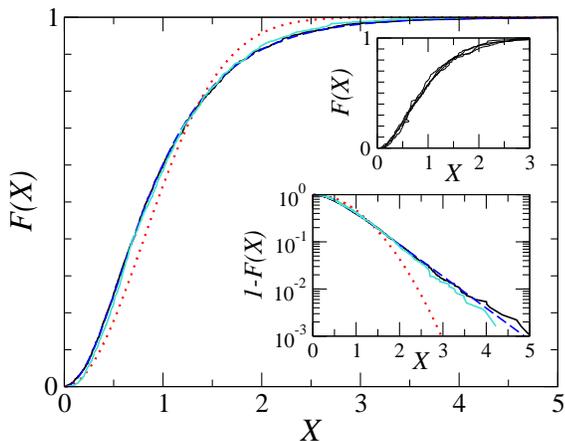}
\caption{\label{fig2bis} We plot the integrated density distribution
$F(X)$ extracted from Internet tournaments (same as Fig.~\ref{fig2};
black line), and from the four main events of the WPT 2006 season
(ranking after day 1 and day 2 totalizing 1256 players; turquoise
line), and compare them to the numerical solution of the exact
Eq.~(\ref{kernel2}) (dashed line), and to the analytical result
Eq.~(\ref{scal1}) of the poker model for $q=0$ (dotted line). The
bottom insert illustrates the exponential decay of $1-F(x)$, instead
of the Gaussian decay predicted by the simple model with $q=0$. In
the top insert, we plot $F(X)$ for each individual WPT tournaments
(day 1 only; for the sake of clarity).}
\end{figure}

In Fig.~\ref{fig2}, we plot the normalized distribution $f(X)$ as a
function of $X=x/\bar x(t)$ obtained from extensive numerical
simulations of the present poker model, with $q$ given by
Eq.~(\ref{q}). We find a perfect data collapse on the numerical
solution of the exact scaling equation Eq.~(\ref{kernel2}). In order
to check the relevance of this parameter-free distribution to real
poker tournaments, we visited two popular on-line poker playing
zones, and followed 20 no-limit Texas hold'em tournaments with an
initial number of players in the range $250-800$. When the number of
players was down to the range $N\sim 60-130$, we manually recorded
their number of chips \cite{disclaim}. Fig.~\ref{fig2} shows the
remarkable agreement between these data and the results of the
present model. The maximum of the distribution corresponds to
players holding around 55\% of the average number of chips per
player. In addition, a player owning twice the average stack per
player $(X=2)$ precedes 90\% of the other players, whereas a player
with half the average stack $(X=1/2)$ precedes only 25\% of the
other players. In Fig.~\ref{fig2bis}, we compare these results to
data collected from the four main events of the World Poker Tour
2006 season \cite{wpt}. Although the level of play is incomparably
better than on typical Internet poker rooms (the buy-in of
10000$\,$\$ or more is also incomparable), the stacks distributions
are very similar and decay exponentially (see the prediction of
Eq.~(\ref{expdecay})), as illustrated in the bottom insert of
Fig.~\ref{fig2bis}. The model without all-in events ($q=0$) would
predict a faster Gaussian decay. The fact that we find similar
results for two very different kinds of poker tournaments certainly
justifies the universal nature of the present theory.

\begin{figure}
\hspace{-0.6cm}\psfig{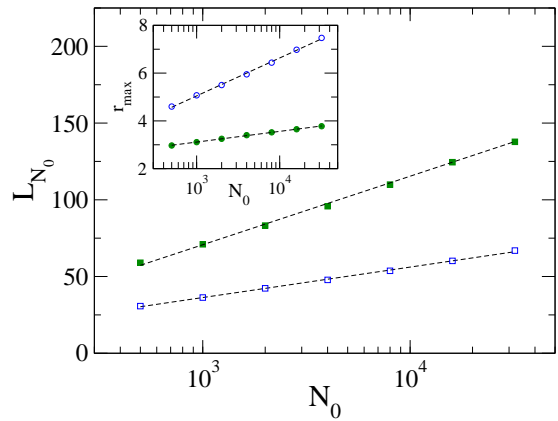}
\caption{\label{fig3} We plot the average number of chip leaders
$L_{N_0}$ as a function of the number of initial players $N_0$,
finding a convincing logarithmic growth
(full symbols correspond to the case $q=0$). The insert shows the
logarithmic growth of $r_{\rm max}$ (defined in the text).
The dashed lines correspond to log-linear fits of the data.}
\end{figure}

\section{Properties of the chip leader}

We now consider the statistical properties of the player with the
largest amount of chips at a given time, dubbed the chip leader.
First, we consider the average number of chip leaders $L_{N_0}$ in a
tournament with $N_0$ initial players. In many competitive
situations \cite{krug,leadcs,krap}, arising for instance in
biological evolution models \cite{krug,leadcs}, it is found that
$L_{N_0}$ grows logarithmically with the number of competing agents
$N_0$, a general result which has been established analytically in
\cite{leadcs}.
\begin{figure}
\psfig{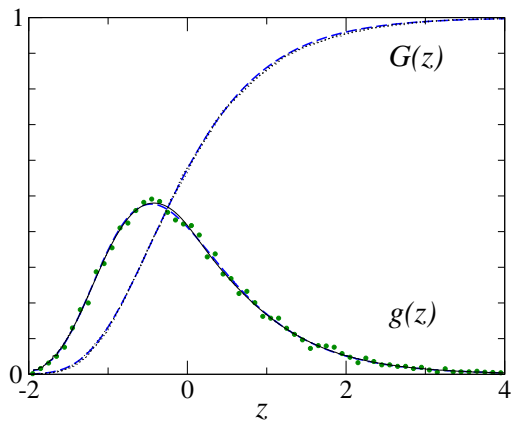} \caption{\label{fig4}
We plot the normalized cumulative distribution $G(z)$ of the fortune
of the chip leader recorded when $N(t)/N_0=40$\%, 20\%, 10\% (thin
dotted lines, all-in processes included, $N_0=10000$, $t_0=2000$,
$x_0/b_0=100$, $10000$ ``tournaments'' played). We also plot the
derivative $g(z)$ of a multi-variable fit of these data (full line)
as well as the standard scatter bin plot (circles). The data
convincingly follow the universal Gumbel distribution (dashed
lines).}
\end{figure}
We confirm that in the present model, with or without all-in
processes, the same phenomenon is observed (see Fig.~\ref{fig3}). We
have also computed the average maximum ratio $r_{\rm max}={\sup_t
\bar x_{\rm lead}/\bar x}$. In the present model, $x_{\rm lead}/\bar
x$ increases rapidly on a scale of order $t_0$, and then decays
(almost linearly with time) to $\sim 1.5$, where it becomes non
self-averaging due to large fluctuations at the end of the
tournament. Fig.~\ref{fig3} illustrates the logarithmic growth of
$r_{\rm max}$ as a function of $N_0$. For $N_0=500$, which is
typical of Internet tournaments, we find $r_{\rm max}\approx 4.6$,
which is fully compatible with a superficial analysis of real data.

Extreme value statistics have recently attracted a lot of attention
from physicists in various contexts \cite{extreme}. In this regard,
we have checked that $z=(x_{\rm lead}-\bar x_{\rm
lead})/({\bar{x^2}_{\rm lead}-\bar x^2_{\rm lead}})^{1/2}$ is
distributed according to the universal Gumbel distribution
\begin{equation}
g(z)=\frac{\pi}{\sqrt{6}}\exp[-Z-\exp(-Z)],
\end{equation}
where $Z=\pi z/\sqrt{6}+\gamma$, and $\gamma$ is Euler's constant.
Such a behavior, which is typical of independent, or at least weakly
correlated random variables \cite{gumbel}, is illustrated on
Fig.~\ref{fig4}.

\section{Conclusion}

In this paper, we have developed a quantitative theory of poker
tournaments and made the connection between this problem and
persistence in physics, the leader problem in evolutionary biology,
and extreme value statistics. In particular, we have identified the
two main ingredients controlling the dynamics of a tournament: the
exponential increase of the blind, and the necessity to include
all-in events where at least two players bet their entire stack. In
order to mimic the play of ``intelligent'' players, we found that
the probability of going all-in should take a well-defined value.
This theory leads to a quantitative understanding of the
scale-invariant stack distribution observed in Internet and WPT
tournaments, and predicts rich statistical features concerning the
chip leader.

In a future work \cite{csnext}, we plan to implement in our model
the optimal strategies for folding, betting or going all-in, hence
eliminating the only free parameter $q$. Preliminary results
\cite{csnext} indicate that the optimal probability $q_0$ to be the
first to go all-in is a simple function of the current pot $P$, of
the chip stack $x$ of the considered player, and of $x_k$, the stack
of the $k$-th player left to bet (among a total of $n$ such
players). Defining $X_k=\min(x_k,x)$, and $q_k$ as the probability
that the player $k$ calls the all-in bet of the first player (and
neglecting multiple calls), we find \cite{csnext}
\begin{eqnarray}
q_k&=&q_0\frac{X_k+P}{X_k+2P},\label{qoptk}\\
P&=&\sum_{k=1}^n q_k X_k\prod_{j=k}^n\left(1-q_j\right)^{-1}
,\label{qopt2}
\end{eqnarray}
where $q_0$ is the solution of the implicit  Eq.~(\ref{qopt2}),
after inserting the expression of $q_k$ obtained in
Eq.~(\ref{qoptk}). A detailed analysis of
Eqs.~(\ref{qoptk},\ref{qopt2}) reveals that the obtained optimal
strategy perfectly reproduces qualitative features observed in real
tournaments, notably the fact that players with a small stack go
more often all-in than others (and are often called). In addition,
direct confrontations between two players owning a big stack (in
units of $\bar x$) are rare, except if the pot is already huge, and
only happens when both players have a very good hand.

Finally, it would be interesting to obtain access to the full
dynamical evolution of a large sample of real-life poker
tournaments, in order to check the predictions of the model
concerning the chip leader and to identify other remarkable
statistical properties of poker tournaments.

\acknowledgments I am very grateful to D.~S. Dean and J. Basson for
fruitful remarks on the manuscript. This work has been exclusively
funded by CNRS and University Paul Sabatier.

\end{document}